\documentclass[letterpaper, 10 pt, conference]{ieeeconf}

\IEEEoverridecommandlockouts  
\usepackage{amsmath,amssymb,amsfonts}
   
\usepackage{amsthm}
\usepackage{booktabs}
\usepackage{tikz}
\usepackage{circuitikz}
\usepackage{algorithm}
\usepackage{algorithmic}

\usetikzlibrary{calc}

\newtheorem{theorem}{Theorem}
\newtheorem{lemma}{Lemma}
\newtheorem{remark}{Remark}
\newtheorem{problem}{Problem}
\newtheorem{definition}{Definition}

\newcommand{\R}{\mathbb{R}}
\newcommand{\wt}[1]{\tilde{#1}}
\newcommand{\wbar}{\bar{\omega}}
\newcommand{\T}{^{\top}}
\newcommand{\Ibat}{I_{\mathrm{bat}}}

\begin{document}

\title{\LARGE \bf
State-Feedback Passivity Control for Synchronous Machine Systems
}

\author{Maryam Khodabakhshloo$^{1}$, Elizabeth L. Ratnam$^{2}$, and Ian R. Petersen$^{1}$
\thanks{$^{1}$M. Khodabakhshloo and I. R. Petersen are with the School of
        Engineering, The Australian National University, Canberra, ACT 2601,
        Australia}%
\thanks{$^{2}$E. L. Ratnam is with the Department of Electrical and Computer
        Systems Engineering, Monash University, Melbourne, VIC 3800,
        Australia}%
}

\maketitle
\thispagestyle{empty}
\pagestyle{empty}

\begin{abstract}
The passivity of a generating station enables network stability to be established from conditions verified locally at each station. For synchronous machines, however, passivity can only be certified under conditions on the operating point, and these conditions fail under heavy loading and weak damping. This paper develops a state-feedback controller that enforces passivity using a storage battery connected at the terminal bus. A bilinear term arising from the product of the rotor-speed and stator-current deviations is treated as a bounded uncertainty, leading to robust positive-real conditions that can be enforced by a state feedback control designed using a single algebraic Riccati equation. The resulting controller guarantees passivity whenever the Riccati equation admits a positive-definite solution, with no separate condition on the operating point. Numerical examples show that existing conditions certify passivity only up to a fraction of the rated current, whereas the proposed state-feedback approach certifies it at the rated operating point of the machine.
\end{abstract}

\section{Introduction}
\noindent Synchronous generators are being replaced by converter-interfaced sources such as wind and solar generation. This changes the conditions under which a power system must be shown to be stable. As the inertia of the system falls, the damping of the remaining machines is smaller, and the machines must operate over a wider range of conditions than before~\cite{Kundur2004,Gu2022}. At the same time, as the number of connected units grows and becomes more diverse, it becomes more difficult to study stability using a single model for the whole network. It is more practical to check a property for each unit separately and then to conclude that the interconnection is stable.

\noindent Classical transient stability analysis relies on assumptions that are difficult to justify in systems with a high penetration of converter-interfaced resources. The swing equation written in terms of power assumes that the rotor speed remains close to its nominal value, which excludes the speed deviations that are central to our passivity analysis~\cite{Monshizadeh2016,Zhou2009}, and the use of phasors assumes that all waveforms remain sinusoidal, which is not guaranteed during a transient~\cite{Caliskan2014}. Reduced models that neglect the stator dynamics can also miss instabilities that appear in more detailed models of the same machine. We therefore use a first-principles model in which the stator currents are states and the electromagnetic torque is retained without approximation, as in~\cite{Natarajan2014,Barabanov2017,Caliskan2014}.

\noindent Passivity theory is well suited to the analysis of such models. A passive system does not generate energy internally, and the interconnection of passive systems is again passive, so a property proved for each unit extends to the whole network. Port-Hamiltonian models give transient stability conditions checked for each generator separately~\cite{Caliskan2014}; the terminal passivity of synchronous generators and converter-interfaced units has been analysed in the direct-quadrature domain~\cite{Dey2021}; and power systems built from Park generator models have been decomposed into subsystems whose equilibrium-independent passivity is established separately~\cite{Nishino2025}. These are analysis results, each providing a sufficient condition whose outcome is fixed by the machine and its operating point. The condition of~\cite{Caliskan2014} requires the damping to be large relative to the steady-state current, making it restrictive under heavy loading or weak damping, while the analysis of~\cite{Nishino2025} holds at every feasible equilibrium but assumes lossless lines and, for the network subsystem, a region where an energy function is convex. When a test fails, the machine may still be passive, but the analysis cannot certify it or change the outcome.

\noindent Passivity has also been enforced by feedback through several actuators. The field voltage and a thyristor-controlled series capacitor are used in~\cite{Sun2009,Li2022} to make a single-machine infinite-bus system feedback passive, while~\cite{Caliskan2014} inserts a fixed series resistance through a static synchronous series compensator, sized offline so that a relaxed condition holds. Such a resistance dissipates power whenever the machine carries current, and the loss grows with the square of that current, so the approach becomes costly for a machine operating near its rated operating point. Energy storage has been used as the actuator in passivity and negative imaginary designs~\cite{Chen2024}, though for a reduced third-order flux-decay model rather than the detailed model considered here.

\noindent In this paper, we consider a synchronous machine connected to a local resistive load and to the grid, with a storage battery at the same bus acting as the control actuator. Since the battery exchanges energy directly with the network, no additional compensating device is required, and no active power is dissipated in steady state. The control objective is to ensure passivity of the generating station at its network interface, where the port variables are the injected grid current and the terminal voltage. The nonlinearity of the perturbed model in direct-quadrature coordinates is a result of angular velocity and current deviations. Instead of neglecting these terms through linearization, we model the angular velocity deviation as a bounded uncertainty parameter, so that the passivity conditions remain valid for all operating trajectories whose angular velocity deviation lies within a prescribed range. Representing the uncertainty by a quadratic inequality with a scaling parameter leads to robust positive-real conditions that can be tested using a single algebraic Riccati equation. This process can be extended to yield a passifying state-feedback gain in closed form. Positive real control based on Riccati equations is well established in the literature; e.g see ~\cite{SunKhargonekarShim1994}. The proposed approach employs state-feedback control, which can be constructed using a single Riccati equation. Whenever this equation has a positive definite solution, the controlled station can be guaranteed to be passive without imposing any condition on the operating point. Passivity therefore becomes a controller design objective.

\noindent The paper is organised as follows. Section~\ref{sec:prelim}
recalls passivity notions and the positive real lemma used
throughout. Section~\ref{sec:model} presents the system model, its
linear fractional representation, and the two problems considered.
Section~\ref{sec:analysis} develops robust passivity analysis and
its connection to a Riccati equation. Section~\ref{sec:design} derives
the passifying state feedback controller, and Section~\ref{sec:numerical} presents numerical examples.

\section{Preliminaries}\label{sec:prelim}
\noindent Dissipativity, introduced by Willems~\cite{Willems1972}, provides a general energy based framework for input-output systems. Passivity is a special case of dissipativity in which the supply rate is the bilinear form $u\T y$.

\begin{definition}\label{def:passivity}
 The system $\dot{x}=f(x,u)$, $y=g(x,u)$, with $x\in\R^{n}$ and $u,y\in\R^{m}$, is \emph{passive} if there exists a continuously differentiable storage function $V:\R^{n}\to\R_{\ge0}$, $V(0)=0$, such that
 \begin{equation}\label{eq:diss-ineq}
 \dot{V}(x)\;\le\; u^{\top}y
 \end{equation}
 along all trajectories.
\end{definition}

\noindent In power systems, the operating point is set by the steady-state power flow, so we apply this notion to the \emph{perturbed dynamics} with respect to a chosen equilibrium $(x^{s},u^{s},y^{s})$. That is, we consider $\tilde{x}:=x-x^{s}$, $\tilde{u}:=u-u^{s}$, $\tilde{y}:=y-y^{s}$, with $V$ measuring energy stored relative to that operating point~\cite{KhodabakhshlooEquilibria2025}.

\noindent The following positive real lemma, also known as the Kalman--Yakubovich--Popov (KYP) Lemma \cite{SunKhargonekarShim1994,AndersonVongpanitlerd1973}, provides necessary and sufficient conditions for a transfer function to be positive real.

\begin{lemma}\label{lem:kyp}
Let $Z(s)=C(sI-A)^{-1}B+D$ be an $m\times m$ real-rational transfer
matrix with $Z(\infty)<\infty$ and minimal realization $\{A,B,C,D\}$. Then $Z(s)$ is positive real if and only if there
exist real matrices $P=P\T\succ 0$, $L\in\R^{n\times m}$ with
$W_0\in\R^{m\times m}$ such that
\begin{subequations}\label{eq:kyp-triple}
\begin{align}
PA+A\T P &= -LL\T, \label{eq:kyp-a}\\
PB - C\T &= -LW_0, \label{eq:kyp-b}\\
D+D\T &= W_0\T W_0. \label{eq:kyp-c}
\end{align}
\end{subequations}
\end{lemma}

\noindent When the feed-through matrix $D+D\T$ is nonsingular, the auxiliary variables $L$ and $W_0$ can be eliminated from~\eqref{eq:kyp-triple}, and the three coupled equations reduce to a single algebraic Riccati equation (ARE) in $P$~\cite[Thm.~6.2.1]{AndersonVongpanitlerd1973}. We use this reduction in the uncertain setting below, both to verify the open-loop system and to design the controller.

\section{System Modeling and Problem Formulation}\label{sec:model}
\noindent Starting from the synchronous machine model derived in~\cite{Grainger1994,KhodabakhshlooEquilibria2025}, we incorporate a load bus, grid connection, the battery energy storage system to obtain the complete station dynamics. We then transform the model from the $abc$ frame to the $dq$ frame, and shift it to a chosen operating point.

\subsection{Synchronous machine with a storage battery}
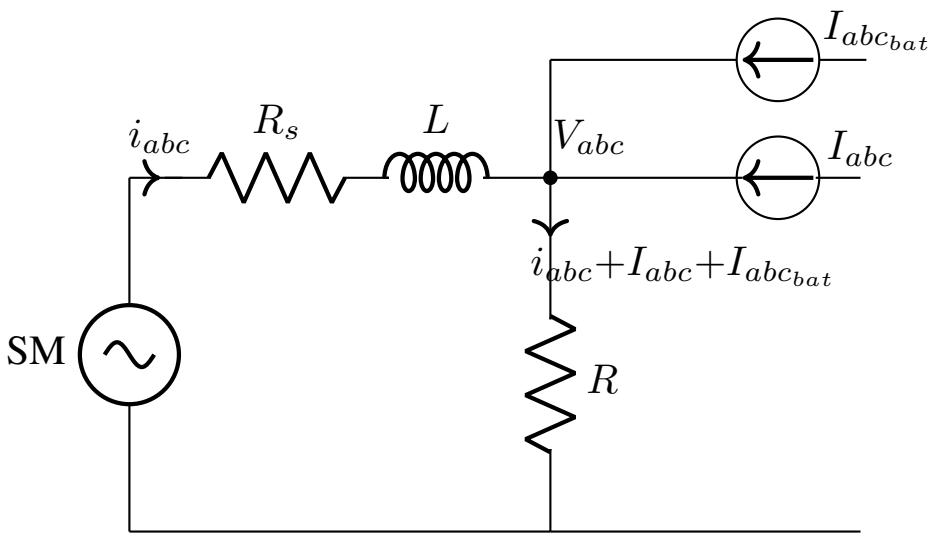
\begin{figure}[t]
\centering
\begin{circuitikz}[scale=0.78, transform shape]
    \draw (0,0) to[sinusoidal voltage source, l_=SM] (0,-3);
    \draw [color=black] (0.3,0) to[R, l=$R_s$](2.2,0) to[L, l=$L$] (3,0);
    \draw (0,0) -- (0.45,0);
    \draw (3,0) -- (3.57,0);
    \draw (3.57,0) -- (5.2,0);
    \draw (5.82,0) -- (6.2,0);
    \draw (3.57,0) -- (3.57,-1.2);
    \draw [color=black] (3.57,-1.2) to[R, l=$R$] (3.57,-2.3);
    \draw (3.57,-2.3) -- (3.57,-3);
    \draw (0,-3) -- (3.57,-3);
    \draw (3.57,-3) -- (6.2,-3);
    \draw (5.5,0) circle (0.35cm);
    \draw (5.5,1) circle (0.35cm);
    \draw[->, thick] (5.8,0) -- (5.2,0);
    \draw[->, thick] (5.8,1) -- (5.2,1);
    \draw[->, thick] (0.25,0) -- (.261,0);
    \draw[->, thick] (3.57,-.45) -- (3.57,-.5);
    \draw (3.57,0) -- (3.57,1);
    \draw (3.57,1) -- (5.15,1);
    \draw (5.85,1) -- (6.25,1);
    \filldraw[black] (3.57,0) circle (1.5pt);
    \node at (6.2,.25) {$I_{abc}$};
    \node at (6.35,1.25) {$I_{abc_{bat}}$};
    \node at (3.9,.35) {$V_{abc}$};
    \node at (.26,.35) {$i_{abc}$};
    \node at (4.7,-.75) {$i_{abc}{+}I_{abc}{+}I_{abc_{bat}}$};
\end{circuitikz}
\caption{Synchronous machine connected to a resistive load and a grid
current source, with a storage battery in parallel at the load bus.}
\label{fig:circuit}
\end{figure}

\noindent We consider a cylindrical-rotor synchronous machine (SM) under the standard assumptions of transient stability analysis~\cite{Barabanov2017}: balanced and sinusoidally distributed stator windings, constant field current $i_f$, negligible damper winding dynamics and magnetic saturation, and a single pole pair. The mechanical dynamics are written in terms of the rotor angle $\theta$ and speed $\omega=\dot\theta$. The electrical dynamics are written in terms of the three phase stator currents $i_{abc}\in\R^{3}$, which under balanced operation satisfy $i_a+i_b+i_c=0$. The machine is connected through its stator resistance and inductance $R_s,L$ to a local bus with resistive load $R$, exchanging power with a grid modeled as a current source $I_{abc}$ (Fig.~\ref{fig:circuit}). A storage battery, modeled as a controllable current source $\Ibat{}_{,abc}$, is added in parallel at the same bus. The grid current can be thought of as an external disturbance input, while the battery current is the control input to be designed.

\noindent Newton's second law applied to the rotor together with Kirchhoff's laws applied to the stator circuit and the load bus give the following station model in the three-phase frame,
\begin{equation}\label{eq:abc}
\begin{aligned}
\dot{\theta}   &= \omega, \\
J\dot{\omega}  &= -D_r \omega
 - b\,\mathbf{sin}(\theta)\T i_{abc} + T_m, \\
L\,\dot{i}_{abc} &= -(R_s+R)\,i_{abc} + \omega b\,\mathbf{sin}(\theta)
 - R\,I_{abc} - R\,\Ibat{}_{,abc},
\end{aligned}
\end{equation}
where $\mathbf{sin}(\theta):=\mathrm{col}\bigl(\sin\theta,\,
\sin(\theta-\tfrac{2\pi}{3}),\,\sin(\theta-\tfrac{4\pi}{3})\bigr)$
represents the rotor position dependence of the stator field coupling,
$J$ is the rotor inertia, $D_r$ the damping coefficient, $T_m$ the
mechanical torque, and $b=\sqrt{\tfrac{3}{2}}M_f i_f>0$ the back-EMF constant, with $M_f$
the peak stator field mutual inductance; a detailed derivation from the flux-linkage equations is given
in~\cite{Grainger1994,KhodabakhshlooEquilibria2025}. Kirchhoff's current law at the bus gives the terminal voltage
$V_{abc}=R\,(i_{abc}+I_{abc}+\Ibat{}_{,abc})$.

\noindent Because the stator field coupling $\mathbf{sin}(\theta)$ rotates with the rotor, the three-phase model~\eqref{eq:abc} is time-varying. We therefore apply the Park transformation $T(\theta)$~\cite{Schiffer2016}, which maps the three phase variables into the two axis representation $x_{dq}=T(\theta)x_{abc}$, with the direct ($d$) axis along the rotor field and the quadrature ($q$) axis $90^{\circ}$ ahead. The transformation is orthogonal on the balanced subspace, $T(\theta)T(\theta)\T=I_2$, so instantaneous power is preserved across the two frames, and we take the rotor angle $\theta$ itself as the reference angle, so that the frame rotates with the rotor and $\dot\theta=\omega$. Under balanced operation the transformed voltages and currents are constant at synchronous steady state. Applying $T(\theta)$ to~\eqref{eq:abc} yields
\begin{equation}\label{eq:dq}
\begin{aligned}
J\dot{\omega} &= -D_r\omega - b\,i_q + T_m, \\
L\,\dot{i}_d  &= -(R_s+R)\,i_d - L\omega i_q - R I_d - R \Ibat{}_{,d}, \\
L\,\dot{i}_q  &= L\omega i_d - (R_s+R)\,i_q +b\omega- R I_q - R \Ibat{}_{,q},
\end{aligned}
\end{equation}
where the rotor angle does not appear: the sinusoidal torque converts to the single term $-b\,i_q$, and the rotating back-EMF reduces to the cross coupling terms $\pm L\omega\, i_{q,d}$ and $b\omega$.

\noindent Let $(\omega^s,i^s_d,i^s_q,I^s_d,I^s_q)$ denote a steady
state of~\eqref{eq:dq}, with $I_{dq_{bat}}^s=0$ since the battery only supplies transient power.
Introducing the deviation variables
$\wt{\omega}$, $\wt{i}_{dq}$, $\wt{I}_{dq}$,
$\wt{I}_{dq_{bat}}$ about this operating point and subtracting
the equilibrium relations from~\eqref{eq:dq} gives the perturbed
dynamics
\begin{equation}\label{eq:perturbed}
\begin{aligned}
J\dot{\wt{\omega}} &= -D_r\wt{\omega} - b\,\wt{i}_q, \\
L\,\dot{\wt{i}}_d  &= -(R_s{+}R)\,\wt{i}_d
   - L\wt{\omega}\,\wt{i}_q - L\wt{\omega}\, i_q^s - L\omega^s\wt{i}_q \\
&\quad\; - R\wt{I}_d - R\wt{I}_{\mathrm{d}_{bat}}, \\
L\,\dot{\wt{i}}_q  &= -(R_s{+}R)\,\wt{i}_q
   + L\wt{\omega}\,\wt{i}_d + L\wt{\omega}\, i_d^s + L\omega^s\wt{i}_d
   + b\wt{\omega} \\
&\quad\; - R\wt{I}_q - R\wt{I}_{\mathrm{q}_{bat}}.
\end{aligned}
\end{equation}

\subsection{Linear Fractional Representation of the Uncertainty}
\noindent The perturbed model~\eqref{eq:perturbed} is nonlinear because of the cross coupling products $\wt{\omega}\,\wt{i}_d$ and $\wt{\omega}\,\wt{i}_q$ between the rotor speed deviation and the stator current deviations. Rather than neglecting these terms, as a linearization about the operating point would do, we treat the speed deviation as an unknown but bounded parameter. In practice, rotor speed deviations remain within specified operating limits, so we assume
\begin{equation}\label{eq:wbar}
|\wt{\omega}(t)|\le\wbar \qquad \forall t\ge0,
\end{equation}
where $\wbar>0$ is a known bound, for example the protection limit on the acceptable frequency deviation. This represents the nonlinear terms as a norm-bounded uncertainty, so that the subsequent results hold for all speed deviations satisfying~\eqref{eq:wbar}.
From the perturbation variables, we define the state, the disturbance, the control input, and the output as
\begin{equation}
x=\begin{bmatrix}\wt{\omega}\\ \wt{i}_d\\ \wt{i}_q\end{bmatrix},\quad
w=\begin{bmatrix}\wt{I}_d\\ \wt{I}_q\end{bmatrix},\quad
u=\begin{bmatrix}\wt{I}_{\mathrm{d}_{bat}}\\ \wt{I}_{\mathrm{q}_{bat}}\end{bmatrix},\quad
y=\begin{bmatrix}\wt{V}_d\\ \wt{V}_q\end{bmatrix},
\end{equation}
where the disturbance $w$ is the grid-current deviation, the control $u$
is the storage battery deviation, and the output $y=R(\wt{i}_{dq}+\wt{I}_{dq}+\wt{I}_{\mathrm{dq}_{bat}})$ is the
deviation of the terminal voltage at the load bus. The pair $(w,y)$ is the port across which passivity is required.

\begin{figure}[!t]
\centering
\begin{tikzpicture}[scale=.75, transform shape, >=latex, thick,
    block/.style={draw, minimum width=1.7cm, minimum height=1.0cm}]
\node[block] (G) at (0,0) {\Large $G$};
\node[block] (K) at (0,-2.3) {\Large $K$};
\node[block] (phi) at (0,2.3) {\Large $\Delta$};
\node[left]  at ($(G.west)+(0,0.65)$) {\Large$z$};
\node[left]  at ($(G.west)+(-1,0)$) {\Large$w$};
\node[left]  at ($(G.west)+(0,-0.65)$) {\Large $u$};
\node[right] at ($(G.east)+(0,0.65)$) {\Large $\zeta$};
\node[right] at ($(G.east)+(1,0)$) {\Large $y$};
\node[right] at ($(G.east)+(0,-0.65)$) {\Large $x$};
\draw[->] (-1.6,0.32) -- ($(G.west)+(0,0.32)$);
\draw[->] (-1.8,0) -- ($(G.west)+(0,0)$);
\draw[->] (-1.6,-0.32) -- ($(G.west)+(0,-0.32)$);
\draw[->] ($(G.east)+(0,0.32)$) -- (1.6,0.32);
\draw[->] ($(G.east)+(0,0)$) -- (2,0);
\draw[->] ($(G.east)+(0,-0.32)$) -- (1.6,-0.32);
\draw[->] (1.6,0.32) |- (phi.east);
\draw[->] (phi.west) -| (-1.6,0.32);
\draw[->] (1.6,-0.32) |- (K.east);
\draw[->] (K.west) -| (-1.6,-0.32);
\end{tikzpicture}
\caption{Linear fractional representation of the station. The scalar
block $\Delta=\wt{\omega}$ represents the nonlinear cross-coupling and
the state feedback $K$ realizes the storage battery. Passivity is
required from $w$ to $y$.}
\label{fig:lft-block}
\end{figure}
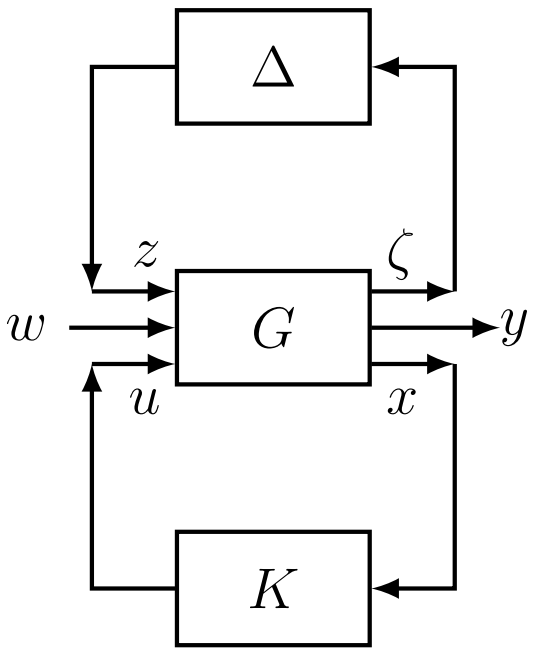

\noindent With these signals, the perturbed dynamics~\eqref{eq:perturbed}
are written as
\begin{equation}\label{eq:lft}
\begin{aligned}
\dot{x} &= Ax + B_1 z + B\,u + B\,w, \qquad z = \Delta\,\zeta, \\
\zeta &= C_1 x, \\
y &= Cx + D\,u + D\,w,
\end{aligned}
\end{equation}
where the scalar uncertain parameter is $\Delta=\wt{\omega}$ with
$|\Delta|\le\wbar$, and
\begin{equation}\label{eq:matrices}
\begin{aligned}
A&=\begin{bmatrix}
-\tfrac{D_r}{J} & 0 & -\tfrac{b}{J}\\[2pt]
-i_q^s & -\tfrac{R_s+R}{L} & -\omega^s\\[2pt]
\tfrac{b+Li_d^s}{L} & \omega^s & -\tfrac{R_s+R}{L}
\end{bmatrix},
\quad
B=\begin{bmatrix}
0 & 0\\ -\tfrac{R}{L} & 0\\ 0 & -\tfrac{R}{L}
\end{bmatrix},\\[4pt]
B_1&=I_3,\qquad
C_1=\begin{bmatrix} 0&0&0\\ 0&0&-1\\ 0&1&0\end{bmatrix},\\[4pt]
C&=\begin{bmatrix} 0&R&0\\ 0&0&R\end{bmatrix},\qquad
D=R\,I_2 .
\end{aligned}
\end{equation}
Eliminating the internal signals $\zeta$ and $z$ from~\eqref{eq:lft} gives the compact form
\begin{equation}\label{eq:lft-compact}
\begin{aligned}
\dot{x} &= \bigl(A+B_1\Delta C_1\bigr)x + B\,u + B\,w,\\
y &= Cx + D\,u + D\,w.
\end{aligned}
\end{equation}
This is the standard linear fractional transformation (LFT) form~\cite{ZhouDoyleGlover1996}, in which the dynamics are split into a linear part $G$ and an uncertain block $\Delta$ closed in feedback around it, as shown in Fig.~\ref{fig:lft-block}. The fixed part $G$ is linear, so passivity of the port $(w,y)$ can be characterized by the KYP conditions of Lemma~\ref{lem:kyp}, while the effect of the feedback block $\Delta$ is handled by a bound on the channel $\zeta \mapsto z$. The controller $K$ acts through the battery on the same port and is designed in Section~\ref{sec:design}.

\subsection{Problem statements}
\noindent A station that is passive from the grid
current $w$ to the terminal voltage $y$ can be connected to
the grid, to transmission lines, or to other passive nodes without
affecting the stability of the resulting network. This property is the
objective of the present paper, and it leads to the following two
problems.

\begin{problem}
\label{prob:analysis}
Consider the uncertain system~\eqref{eq:lft-compact} with $u\equiv0$.
Determine whether there exists $P=P\T\succ0$ such that the storage
function $V(x)=\frac{1}{2}x\T Px$ satisfies
\begin{equation}\label{eq:diss-open}
\dot{V}(x)\;\le\; w\T y
\end{equation}
for all admissible uncertainties $|\Delta|\le\wbar$.
\end{problem}

\begin{problem}
\label{prob:design}
Consider the uncertain system~\eqref{eq:lft-compact} and assume that no
such $P$ exists. Design a state feedback $u=Kx$, realized by the
storage battery, such that the closed-loop system
\begin{equation}\label{eq:cl}
\begin{aligned}
\dot{x} &= \bigl(A_{\mathrm{cl}}+B_1\Delta C_1\bigr)x + B w,\\
y &= C_{\mathrm{cl}}\,x + Dw,
\end{aligned}
\qquad
\begin{aligned}
A_{\mathrm{cl}} &:= A+BK,\\
C_{\mathrm{cl}} &:= C+DK,
\end{aligned}
\end{equation}
is passive from $w$ to $y$ for all $|\Delta|\le\wbar$.
\end{problem}

\section{A Riccati Equation  Approach to Robust Passivity Analysis}
\label{sec:analysis}
\noindent The KYP conditions of Lemma~\ref{lem:kyp} apply to a linear system, whereas the state matrix of~\eqref{eq:lft-compact} depends on the unknown parameter $\Delta$. We therefore bound the uncertainty with a scaling parameter $\epsilon>0$, obtaining KYP conditions that certify passivity for every $|\Delta|\le\wbar$, and reduce them to a single algebraic Riccati equation in $P$.

\subsection{Bounding the uncertainty}
\noindent The constraint $|\Delta|\le\wbar$ is quadratic, and the S-procedure handles it by adding a nonnegative multiple of it to the Lyapunov inequality~\cite{Boyd1994}. For the scalar uncertainty considered here, this reduces to the following bound.
\begin{lemma}\label{lem:sproc}
Let $P=P\T$, let $B_1=I_3$ and $C_1$ be as in \eqref{eq:matrices}, and
let $\Delta\in\R$ with $|\Delta|\le\wbar$. Then, for every
$\epsilon>0$ and all $x\in\R^3$,
\begin{multline}\label{eq:sproc}
x\T\bigl(PB_1\Delta C_1+C_1\T\Delta B_1\T P\bigr)x\\
\le \epsilon^2\, x\T B_1B_1\T x
+\frac{\wbar^2}{\epsilon^2}\, x\T P C_1\T C_1 P\, x .
\end{multline}
\end{lemma}

\begin{proof}
Since $\Delta$ is scalar and $B_1=I_3$, the left hand side equals
$2\Delta\, x\T P C_1 x = 2\Delta\,\langle x,\, C_1\T P x\rangle$, using the fact
that a scalar equals its transpose. By Young's inequality,
$2\Delta\langle x, C_1\T Px\rangle
\le \epsilon^2\|x\|^2+\epsilon^{-2}\Delta^2\|C_1\T Px\|^2
\le \epsilon^2 x\T x+\tfrac{\wbar^2}{\epsilon^2}x\T PC_1C_1\T Px$.
The claim follows since $B_1B_1\T=I_3$ and, for the skew-symmetric $C_1$
in \eqref{eq:matrices}, $C_1C_1\T=C_1\T C_1=\mathrm{diag}(0,1,1)$.
\end{proof}
\noindent The scalar $\epsilon>0$ is a free scaling parameter and the only design parameter in the subsequent analysis.

\subsection{Robust KYP conditions}
\noindent The bound~\eqref{eq:sproc} replaces the uncertain term
$PB_1\Delta C_1 + C_1\T\Delta B_1\T P$ by two terms that no longer
contain $\Delta$. Adding them to the first KYP
equation~\eqref{eq:kyp-a} gives conditions that are
independent of the uncertainty, but strong enough to imply passivity for
every $|\Delta|\le\wbar$.
\begin{theorem}
\label{thm:robust-kyp}
Suppose there exist $\epsilon>0$, $P=P\T\succ0$, $L\in\R^{3\times2}$ and
$W_0\in\R^{2\times2}$ such that
\begin{subequations}\label{eq:robust-kyp}
\begin{align}
PA+A\T P+\epsilon^2 B_1B_1\T+\frac{\wbar^2}{\epsilon^2}PC_1\T C_1P
 &= -LL\T, \label{eq:rk-a}\\
PB-C\T &= -LW_0, \label{eq:rk-b}\\
D+D\T &= W_0\T W_0. \label{eq:rk-c}
\end{align}
\end{subequations}
Then the uncertain system~\eqref{eq:lft-compact} with $u\equiv0$ is
passive from $w$ to $y$, with storage function
$V(x)=\tfrac12\,x\T Px$, for every $|\Delta(t)|\le\wbar$.
\end{theorem}

\begin{proof}
Along the trajectories of~\eqref{eq:lft-compact} with $u\equiv0$, the
derivative of $V(x)=\tfrac12 x\T Px$ is
\begin{equation}\label{eq:Vdot}
\dot{V} = x\T P\dot{x}
= \tfrac12\, x\T\!\left[P\bar{A}(\Delta)+\bar{A}(\Delta)\T P\right]\!x
 + x\T PBw ,
\end{equation}
where $\bar{A}(\Delta):=A+B_1\Delta C_1$. Expanding the bracket
in~\eqref{eq:Vdot},
\begin{equation}\label{eq:split}
P\bar{A}(\Delta)+\bar{A}(\Delta)\T P
= PA+A\T P + PB_1\Delta C_1 + C_1\T\Delta B_1\T P ,
\end{equation}
and bounding the last two terms by Lemma~\ref{lem:sproc},
\begin{multline}\label{eq:bounded}
x^\top\!\left[P\bar{A}(\Delta)+\bar{A}(\Delta)^\top P\right]\!x \\
\le
x^\top\!\left[
PA+A^\top P
+\epsilon^2 B_1B_1^\top
+\frac{\bar{\omega}^2}{\epsilon^2}PC_1^\top C_1P
\right]\!x\\
= -x^\top LL^\top x ,
\end{multline}
where the equality is~\eqref{eq:rk-a}, so the uncertainty has been
eliminated. For the remaining term in~\eqref{eq:Vdot}, \eqref{eq:rk-b}
gives $PB=C\T-LW_0$, hence
\begin{equation}\label{eq:cross}
x\T PBw = x\T C\T w - x\T LW_0 w .
\end{equation}
The supply rate is, using $y=Cx+Dw$ and the fact that
$w\T Dw$ is a scalar together with~\eqref{eq:rk-c},
\begin{equation}\label{eq:supply}
w\T y = w\T Cx + w\T Dw
= x\T C\T w + \tfrac12\, w\T W_0\T W_0 w .
\end{equation}
Substituting~\eqref{eq:bounded} and~\eqref{eq:cross} into~\eqref{eq:Vdot}
and subtracting~\eqref{eq:supply}, the terms $x\T C\T w$ cancel and
\begin{multline}\label{eq:square}
\dot{V}-w\T y
\;\le\; -\tfrac12\, x\T LL\T x - x\T LW_0 w - \tfrac12\, w\T W_0\T W_0 w\\
\;=\; -\tfrac12\left\|L\T x + W_0 w\right\|^{2} .
\end{multline}
The right-hand side is nonpositive, so $\dot{V}\le w\T y$ for every $\Delta$ with $|\Delta|\le\wbar$, which is the passivity inequality~\eqref{eq:diss-ineq}.
\end{proof}

\begin{remark}
Condition \eqref{eq:rk-a} contains the quadratic term $PC_1\T C_1P$, so \eqref{eq:robust-kyp} is not a linear matrix inequality in $P$. We therefore use the Riccati method below rather than an LMI formulation.
\end{remark}

\subsection{The robust positive-real Riccati equation}
\noindent Among the three unknowns of Theorem~\ref{thm:robust-kyp}, only $P$ is of interest, as it defines the storage function; $L$ and $W_0$ are auxiliary variables introduced to complete the square in~\eqref{eq:square}. Since the feed-through matrix satisfies $M:=D+D\T=2R\,I_2\succ0$, it is nonsingular, and the following result eliminates the auxiliary variables and yields a single quadratic equation in $P$.

\begin{theorem}\label{thm:ol-are}
Suppose $M=D+D\T$ is nonsingular and let $\epsilon>0$ be given. If $P=P\T\succ0$
satisfies
\begin{multline}\label{eq:ol-riccati}
PA+A\T P+\epsilon^2 B_1B_1\T+\frac{\wbar^2}{\epsilon^2}\,PC_1\T C_1P\\
+\,(PB-C\T)\,M^{-1}(B\T P-C)=0,
\end{multline}
then, with $V$ any real orthogonal matrix,
\begin{equation}\label{eq:ol-LW}
W_0=V M^{1/2},\qquad
L=(C\T-PB)\,M^{-1/2}V\T,
\end{equation}
the triple $\{P,L,W_0\}$ satisfies the robust KYP
conditions~\eqref{eq:robust-kyp}. Consequently, the uncertain
system~\eqref{eq:lft-compact} with $u\equiv0$ is passive from $w$ to $y$
for all $|\Delta|\le\wbar$.
\end{theorem}

\begin{proof}
We verify the three conditions~\eqref{eq:robust-kyp} in turn. Since $V$
is orthogonal, $V\T V=I$, and $M^{-1/2}M^{1/2}=I$.
For~\eqref{eq:rk-c}, substituting~\eqref{eq:ol-LW} gives
$W_0\T W_0 = M^{1/2}V\T V M^{1/2} = M = D+D\T$. For~\eqref{eq:rk-b},
\[
-LW_0 = -(C\T-PB)M^{-1/2}V\T V M^{1/2}
= PB-C\T .
\]
For~\eqref{eq:rk-a}, the same substitution and the symmetry of
$M^{-1/2}$ give
\[
\begin{aligned}
LL\T &= (C\T-PB)M^{-1/2}V\T V M^{-1/2}(C-B\T P)\\
     &= (PB-C\T)M^{-1}(B\T P-C),
\end{aligned}
\]
and~\eqref{eq:ol-riccati} states that this equals
$-\bigl[PA+A\T P+\epsilon^2 B_1B_1\T
+\tfrac{\wbar^2}{\epsilon^2}PC_1\T C_1P\bigr]$,
which is~\eqref{eq:rk-a}. Hence, all three conditions of
Theorem~\ref{thm:robust-kyp} hold and the system is passive.
\end{proof}

\noindent The matrix $V$ is arbitrary, since $L$ and $W_0$ are not
unique: if $\{P,L,W_0\}$ satisfies~\eqref{eq:robust-kyp}, so does
$\{P,LV\T,VW_0\}$ for any orthogonal $V$. The storage function $P$ does not
depend on this choice, because it is fixed by~\eqref{eq:ol-riccati}.

\noindent Rearranging the terms in~\eqref{eq:ol-riccati} according to their dependence on $P$ yields the standard algebraic Riccati equation
\begin{equation}\label{eq:ARE-std}
P\widetilde{A}+\widetilde{A}\T P+P\widetilde{Q}P+\widetilde{S}=0,
\qquad P\succ0,\quad \epsilon>0,
\end{equation}
where, for our system~\eqref{eq:matrices} with $M=2R\,I_2$,
\begin{equation}\label{eq:ARE-data}
\begin{aligned}
\widetilde{A}&=A-BM^{-1}C,\\
\widetilde{Q}&=BM^{-1}B\T+\frac{\wbar^2}{\epsilon^2}\,C_1\T C_1\\
&=\mathrm{diag}\Bigl(0,\;
\tfrac{R}{2L^2}+\tfrac{\wbar^2}{\epsilon^2},\;
\tfrac{R}{2L^2}+\tfrac{\wbar^2}{\epsilon^2}\Bigr),\\
\widetilde{S}&=C\T M^{-1}C+\epsilon^2 B_1B_1\T
=\mathrm{diag}\Bigl(\epsilon^2,\;
\tfrac{R}{2}+\epsilon^2,\;\tfrac{R}{2}+\epsilon^2\Bigr).
\end{aligned}
\end{equation}

\subsection{Solving the Riccati equation}
\noindent Equation~\eqref{eq:ARE-std} is quadratic in $P$. For a fixed
value of $\epsilon>0$, we solve it using the invariant subspace method
\cite{AndersonVongpanitlerd1973,ZhouDoyleGlover1996}. We first construct
the Hamiltonian matrix
\begin{equation}\label{eq:hamiltonian}
H(\epsilon)=
\begin{bmatrix}
\widetilde{A} & \widetilde{Q}\\
-\widetilde{S} & -\widetilde{A}\T
\end{bmatrix}\in\R^{6\times6},
\end{equation}
whose spectrum is symmetric about the imaginary axis. If no eigenvalue lies on the imaginary axis, we compute the invariant subspace associated with the three eigenvalues in the open left half plane. Let the corresponding basis be $\mathrm{col}(X_1,X_2)$. If $X_1$ is invertible, we obtain
\begin{equation}\label{eq:P-from-H}
P=X_2X_1^{-1},
\end{equation}
the stabilizing solution of~\eqref{eq:ARE-std}, that is, $\widetilde{A}+\widetilde{Q}P$ is Hurwitz~\cite[Sec.~6.4]{AndersonVongpanitlerd1973}. Problem~\ref{prob:analysis} therefore reduces to finding a single scalar $\epsilon>0$ where the system~\eqref{eq:perturbed} is robustly passive if there exists an $\epsilon$ for which the solution in~\eqref{eq:P-from-H} satisfies $P\succ0$.

\section{Robust Passifying State-Feedback Design}\label{sec:design}
\noindent When the analysis of Section~\ref{sec:analysis} returns no solution, the system~\eqref{eq:lft-compact} is not certified passive at the given operating point, and passivity must be imposed rather than verified. This is Problem~\ref{prob:design}; find a gain $K$ such that the state feedback $u=Kx$, delivered by the storage battery, makes the closed loop satisfy the conditions of Theorem~\ref{thm:robust-kyp}. Written for the closed-loop matrices $A_{\mathrm{cl}}=A+BK$ and $C_{\mathrm{cl}}=C+DK$, the requirement~\eqref{eq:ol-riccati} becomes
\begin{multline}\label{eq:K-equation}
P(A+BK)+(A+BK)\T P
+\epsilon^2 B_1B_1\T+\frac{\wbar^2}{\epsilon^2}PC_1\T C_1P\\
+\bigl(PB-(C+DK)\T\bigr)M^{-1}\bigl(B\T P-(C+DK)\bigr)=0 .
\end{multline}
Both $P$ and $K$ are unknown in~\eqref{eq:K-equation}. The gain,
however, appears only in a quadratic form that can be isolated by completion of squares. By choosing $K$ to cancel the squared term, we obtain an equation involving only $P$.

\subsection{The optimal gain}
\noindent The gain enters~\eqref{eq:K-equation} through terms of degree
zero, one and two. To separate them, we use the definitions
\begin{equation}\label{eq:abn}
\alpha:=PB-C\T,\;\;
\beta:=PB-\alpha M^{-1}D,\;\;
N:=D\T M^{-1}D ,
\end{equation}
with which the last term of~\eqref{eq:K-equation} expands to
\begin{multline}\label{eq:expanded}
PA+A\T P+\epsilon^2B_1B_1\T
+\tfrac{\wbar^2}{\epsilon^2}PC_1\T C_1P
+\alpha M^{-1}\alpha\T\\
+\;\beta K+K\T\beta\T\;+\;K\T NK\;=\;0 .
\end{multline}
Since $D=R\,I_2$, the matrix $N=\tfrac{R}{2}I_2$ is positive definite,
and the $K$-dependent terms can be completed into a square,
\begin{multline}\label{eq:complete-square}
K\T NK+\beta K+K\T\beta\T\\
=\bigl(K+N^{-1}\beta\T\bigr)\T N\bigl(K+N^{-1}\beta\T\bigr)
-\beta N^{-1}\beta\T .
\end{multline}
The first term on the right of~\eqref{eq:complete-square} is positive semidefinite and, because
$N\succ0$, it vanishes if
\begin{equation}\label{eq:K-abstract}
K=-N^{-1}\beta\T,
\end{equation}
which removes $K$ from~\eqref{eq:expanded}. Since $D=R\,I_2$, we have $M=2R\,I_2$,
$M^{-1}D=\tfrac12 I_2$ and $N^{-1}=\tfrac{2}{R}I_2$, so
$\beta=\tfrac12\bigl(PB+C\T\bigr)$ by~\eqref{eq:abn}, and
\begin{equation}\label{eq:K-formula}
K=-\frac{1}{R}\bigl(B\T P+C\bigr).
\end{equation}
The gain therefore depends on the single unknown $P$.

\subsection{The closed-loop Riccati equation}
\noindent Substituting~\eqref{eq:K-abstract} into~\eqref{eq:complete-square} and then into~\eqref{eq:expanded} leaves the $K$-independent terms together with $-\beta N^{-1}\beta\T$. The two quadratic terms combine as
\begin{equation}\label{eq:cancellation}
\alpha M^{-1}\alpha\T-\beta N^{-1}\beta\T
=-\frac{1}{R}\bigl(PBC+C\T B\T P\bigr),
\end{equation}
where the products $PBB\T P$ and $C\T C$ occur in both terms with
opposite signs and cancel. What remains is an equation in $P$ alone.

\begin{theorem}\label{thm:design}
Let $\epsilon>0$ and suppose $P=P\T\succ0$ satisfies
\begin{multline}\label{eq:cl-riccati}
PA+A\T P+\epsilon^2B_1B_1\T
+\frac{\wbar^2}{\epsilon^2}PC_1\T C_1P\\
-\frac{1}{R}\bigl(PBC+C\T B\T P\bigr)=0 .
\end{multline}
Then the state feedback $u=Kx$ with $K$ given by~\eqref{eq:K-formula}
makes the closed-loop system~\eqref{eq:cl} passive from $w$ to $y$,
with storage function $V(x)=\tfrac12 x\T Px$, for every
$|\Delta(t)|\le\wbar$.
\end{theorem}
\begin{proof}
Choosing $K$ according to~\eqref{eq:K-abstract} eliminates the square in~\eqref{eq:complete-square}, so~\eqref{eq:cl-riccati} is equivalent to~\eqref{eq:K-equation}. The latter is exactly~\eqref{eq:ol-riccati} written for the closed-loop realization $\{A_{\mathrm{cl}},B,C_{\mathrm{cl}},D\}$. Applying Theorem~\ref{thm:ol-are} to this realization yields matrices $L$ and $W_0$ that satisfy the robust KYP conditions~\eqref{eq:robust-kyp}, and Theorem~\ref{thm:robust-kyp} then proves passivity of the closed-loop system.
\end{proof}

\noindent The closed loop Riccati equation~\eqref{eq:cl-riccati} has the same standard form as~\eqref{eq:ARE-std},
\begin{equation}\label{eq:cl-ARE-std}
P\widetilde{A}_c+\widetilde{A}_c\T P+P\widetilde{Q}_cP+\widetilde{S}_c=0,
\end{equation}
where
\begin{equation}\label{eq:cl-ARE-data}
\widetilde{A}_c=A-\frac{1}{R}BC,\qquad
\widetilde{Q}_c=\frac{\wbar^2}{\epsilon^2}C_1\T C_1,\qquad
\widetilde{S}_c=\epsilon^2 I_3,
\end{equation}
and is solved by the same Hamiltonian
method~\eqref{eq:hamiltonian}--\eqref{eq:P-from-H}.

\section{Numerical Examples}\label{sec:numerical}
\noindent In earlier work~\cite{Khodabakhshloo2025} we showed that the shifted dynamics of the synchronous machine with local resistive load and no battery is passive from current to voltage if
\begin{equation}\label{eq:shifted-condition}
(i_d^s)^2+(i_q^s)^2\;\le\;\frac{4 D_r R_s}{L^2}.
\end{equation}
This condition is obtained from a single fixed quadratic storage function and requires the mechanical damping to be large relative to the operating current, so it becomes restrictive under heavy loading or weak damping. The open-loop test~\eqref{eq:ARE-std} searches instead over all quadratic storage functions and over the scaling $\epsilon$, so it certifies a larger set of operating points; every point satisfying~\eqref{eq:shifted-condition} satisfies it, but not conversely. When that test also fails, the closed-loop equation~\eqref{eq:cl-ARE-std} may still admit $P\succ0$, in which case the feedback~\eqref{eq:K-formula} makes the station passive with no condition on the operating point, and the standard interconnection results for passive systems apply to the controlled station. We now compare the three tests on a benchmark machine.

\subsection{The design procedure}
\noindent For a given speed deviation bound $\wbar$, each $\epsilon$ yields a closed-loop Riccati equation~\eqref{eq:cl-ARE-std}, which we solve by the Hamiltonian method. The equation is feasible when $H(\epsilon)$ in~\eqref{eq:hamiltonian} has no eigenvalue on the imaginary axis and the solution $P$ from~\eqref{eq:P-from-H} is symmetric and positive definite; otherwise it is discarded. Algorithm~\ref{alg:design} finds the largest speed bound $\wbar_{\max}$ for which the design is feasible, so that it can be compared with the required $\wbar$ (Phase~1); scans $\epsilon$ and keeps the value minimizing the control effort $\|K\|$ (Phase~2); and verifies the design (Phase~3). A second criterion is needed in Phase~2 because every feasible $\epsilon$ certifies passivity; the closed-loop decay rate is not useful here, since the first row of $B$ is zero, so $e_1\T(A+BK)=e_1\T A$ and the slowest mode remains close to $-D_r/J$ for every $\epsilon$.

\begin{algorithm}[t]
\caption{Robust passifying battery feedback.}
\label{alg:design}
\begin{algorithmic}[1]
\REQUIRE physical parameters; speed bound $\wbar$
\ENSURE gain $K$ and storage matrix $P\succ0$, or
infeasibility
\STATE build $A,B,B_1,C_1,C,D$ from~\eqref{eq:matrices}; set
$\widetilde{A}_c=A-\tfrac{1}{R}BC$
\STATE \textbf{Phase 1: largest feasible bound on the speed deviation}
\STATE find, by bisection on $\wbar'$, the largest $\wbar_{\max}$ for
which some $\epsilon\in\mathcal{E}$ solves~\eqref{eq:cl-ARE-std} with
$P\succ0$
\IF{$\wbar>\wbar_{\max}$}
  \STATE \textbf{return} infeasible; reduce $\wbar$
\ENDIF
\STATE \textbf{Phase 2: select the scaling parameter $\epsilon$}
\FOR{each $\epsilon\in\mathcal{E}$}
\STATE solve~\eqref{eq:cl-ARE-std} by the Hamiltonian
  method~\eqref{eq:hamiltonian}--\eqref{eq:P-from-H}
  \IF{$P\succ0$}
    \STATE $K_\epsilon\gets-\tfrac{1}{R}(B\T P+C)$,\quad
    $\eta(\epsilon)\gets\|K_\epsilon\|$
  \ENDIF
\ENDFOR
\STATE $\epsilon^\star\gets\arg\min_{\epsilon}\eta(\epsilon)$
\STATE \textbf{Phase 3: final controller and verification}
\STATE solve~\eqref{eq:cl-ARE-std} at $\epsilon^\star$ for $P$;\quad
$K\gets-\tfrac{1}{R}(B\T P+C)$
\STATE verify $P\succ0$, Riccati residual, $A+BK$ Hurwitz,
Theorem~\ref{thm:robust-kyp} at $\Delta=|\wbar|$
\STATE \textbf{return} $K,\,P$
\end{algorithmic}
\end{algorithm}

\subsection{Operating point and examples}
\noindent The machine is the $160$~MVA, $15$~kV, $60$~Hz synchronous generator of~\cite[Example~4.1]{Anderson2002}, supplying a local load that absorbs $10\%$ of the delivered active power, with the remaining power delivered to the grid; this determines the load resistance $R$. The parameters, listed in Table~\ref{tab:params}, are in physical units. The back-EMF constant follows from the mutual inductance and the field current of~\cite[Example~4.1]{Anderson2002} as $b=\sqrt{3/2}M_F\,i_f$, and the inertia from the inertia constant $H=2.37$~s as $J=2HS/(\omega^s)^2$. The damping coefficient is the nominal value of~\cite{Barabanov2017}, which in physical units gives $D_r=2252$~N$\cdot$m$\cdot$s/rad, and the series impedance comprises the stator of~\cite[Example~4.1]{Anderson2002} together with the connection to the load bus assumed in~\cite{Barabanov2017}. An operating point is then determined by the loading, with the terminal voltage fixed at its rated value and the power factor fixed at $0.85$, so each loading level gives a complete operating point $(\omega^s,i^s_{dq},I^s_{dq},R)$.

\begin{table}[!t]
\centering
\caption{Machine parameters, from~\cite[Example~4.1]{Anderson2002}
and~\cite{Barabanov2017}.}
\label{tab:params}
\begin{tabular}{lll}
\toprule
\textbf{Symbol} & \textbf{Description} & \textbf{Value}\\
\midrule
$S$         & Rated apparent power        & $160$ MVA\\
$V_{LL}$    & Rated line-to-line voltage  & $15$ kV\\
$I_N$       & Rated stator current        & $6158$ A\\
$\omega^s$  & Synchronous speed           & $377$ rad/s\\
$R_s$       & Stator and line  resistance           & $29.7$ m$\Omega$\\
$L$         & Stator and line inductance           & $7.833$ mH\\
$J$         & Rotor inertia               & $5336$ kg$\cdot$m$^2$\\
$D_r$       & Damping coefficient         & $2252$ N$\cdot$m$\cdot$s/rad\\
$\sqrt{3/2}M_f$         & Mutual inductance in $dq$          & $109.01$ mH\\
$\wbar$     & Speed bound                 & $3.14$ rad/s\\
\bottomrule
\end{tabular}
\end{table}

\noindent In the closed-form condition~\eqref{eq:shifted-condition},
two parameters are decisive: the loading
$\|i^s_{dq}\|$ and the damping $D_r$. We first
vary the loading while keeping the damping at its nominal value, and
then reduce the damping.

\textit{Example 1 (a passive operating point):}
At $15\%$ of rated current, $\|i^s_{dq}\|=1600$~A with
$i_d^s=-1124$~A and $i_q^s=1139$~A, the machine delivers $20.4$~MW of a
total $24.0$~MVA, and the local load absorbs $2.04$~MW through
$R=110.3\ \Omega$. Condition~\eqref{eq:shifted-condition} holds and
the open-loop test~\eqref{eq:ARE-std} also certifies passivity, so no
controller is required.

\textit{Example 2 (the closed-form condition fails, but the Riccati test
certifies passivity):} At $30\%$ of rated current, $\|i^s_{dq}\|=3200$~A with
$i_d^s=-2572$~A and $i_q^s=1904$~A, the machine delivers $40.8$~MW of
$48.0$~MVA through $R=55.1\ \Omega$. Since $3200>2087$~A,
condition~\eqref{eq:shifted-condition} is violated. The open-loop
test~\eqref{eq:ARE-std} nevertheless returns a solution $P\succ0$, so
the machine remains passive. This illustrates the conservatism of
condition~\eqref{eq:shifted-condition}, which uses one fixed
storage function, whereas the Riccati test searches over all quadratic
storage functions. Varying the loading shows that the open-loop test certifies passivity up to about half of rated current.

\textit{Example 3 (an operating point not certified by the Riccati test):}
At rated current, $\|i^s_{dq}\|=10\,667$~A with $i_d^s=-9332$~A and
$i_q^s=5167$~A, the machine delivers its rated $160$~MVA, that is
$136$~MW at power factor $0.85$, through $R=16.5\ \Omega$. Neither the
closed-form condition~\eqref{eq:shifted-condition} nor the open-loop Riccati test~\eqref{eq:ARE-std} certifies passivity.
The controller design is feasible with
$\epsilon^\star=0.0225$ and gives the positive definite storage matrix
\begin{equation}\label{eq:P-ex3}
P=
\begin{bmatrix}
2.158\times10^{-1} & -5.81\times10^{-5} & 1.123\times10^{-3}\\
-5.81\times10^{-5} & 8.199\times10^{-5} & 0\\
1.123\times10^{-3} & 0 & 8.199\times10^{-5}
\end{bmatrix},
\end{equation}
and the state feedback gain
\begin{equation}\label{eq:K-ex3}
K=
\begin{bmatrix}
-0.007 & -0.990 & \phantom{-}0.000\\
\phantom{-}0.143 & \phantom{-}0.000 & -0.990
\end{bmatrix},
\qquad\|K\|=1.41.
\end{equation}
The resulting closed-loop poles are $\{-0.24,\,-26.0\pm377j\}$, and Phase~1 of Algorithm~\ref{alg:design} returns $\wbar_{\max}=3.88$~rad/s at this operating point, so the design is feasible for frequency deviations up to $\pm0.617$~Hz. The battery feedback therefore provides a passivity certificate at the rated operating point, where neither analysis does.

\textit{Example 4 (weak damping without a passivity guarantee):}
We now reduce the damping to a quarter of its nominal value,
$D_r=563$~N$\cdot$m$\cdot$s/rad, and return to the operating point of
Example~2. The threshold in~\eqref{eq:shifted-condition} falls from $2087$~A to $1043$~A, and the open-loop test~\eqref{eq:ARE-std} now fails where it succeeded before. The state feedback design remains feasible
with $\epsilon^\star=0.0230$ and
\begin{equation}\label{eq:K-ex4}
K=
\begin{bmatrix}
0.138 & -0.989 & \phantom{-}0.000\\
0.058 & \phantom{-}0.000 & -0.989
\end{bmatrix}.
\end{equation}
Examples~2 and~4 share the same operating point and differ only in the damping. With weaker damping, passivity is no longer guaranteed at the same loading level. As synchronous generators are replaced by converter-interfaced sources, such weakly damped operating conditions are expected to become more common, making battery-based damping increasingly important.

\section{Conclusion}
\noindent We have presented a state-feedback controller, realised by a storage battery at the terminal bus, that passifies a synchronous machine from the grid current to the terminal voltage. Treating the nonlinear coupling between the rotor speed and the stator currents as a norm-bounded uncertainty, the certificate holds for every speed deviation within a prescribed bound rather than at a single operating point, and the conditions reduce to one algebraic Riccati equation from which the gain follows in closed form. In the numerical study the certified range extended to the rated operating point of the machine.

\noindent Extending the design to multi-machine networks is a next step, where each generator is passified locally and passivity of the overall network follows from the interconnection results. The battery is also modelled here as an ideal current source; accounting for its power and energy limits, together with converter saturation, is an important step toward practical implementation.


\end{document}